# Dispersion Measurement of Ultra-High Numerical Aperture Fibers covering Thulium, Holmium, and Erbium Emission Wavelengths

Piotr Ciąćka,[1,2,*] Anupamaa Rampur,[1,3] Alexander Heidt,[4] Thomas Feurer,[4] and Mariusz Klimczak[1]

[1]*Institute of Electronic Materials Technology, Wólczyńska 133, 01-919 Warsaw, Poland*
[2]*Institute of Physical Chemistry, Polish Academy of Sciences, Kasprzaka 44/52, 01-224 Warsaw, Poland*
[3]*Faculty of Physics, University of Warsaw, Pasteura 5, 02-093 Warsaw, Poland*
[4]*Institute of Applied Physics, University of Bern, Sidlerstrasse 5, 3012 Bern, Switzerland*
*\*Corresponding author: pciacka@ichf.edu.pl*

**We present broadband group velocity dispersion (GVD) measurements of commercially available ultra-high numerical aperture fibers (UHNA1, UHNA3, UHNA4, UHNA7 and PM2000D from Coherent-Nufern). Although these fibers are attractive for dispersion management in ultrafast fiber laser systems in the 2 μm wavelength region, experimental dispersion data in literature is scarce and inconsistent. Here we demonstrate the measurements using the spectral interferometry technique covering the typically used erbium, thulium and holmium emission bands. The results are characterized in terms of the standard-deviation uncertainty and compared with previous literature reports. Fitting parameters are provided for each fiber allowing for the straightforward replication of the measured dispersion profiles. This work is intended to facilitate the design of ultrafast fiber laser sources and the investigations of nonlinear optical phenomena.**

## 1. INTRODUCTION

Dispersion management is vital in many facets of ultrafast technology. Fiber oscillators, whether operating in solitonic [1,2], dissipative soliton [3], stretched pulse [4–6], self-similar [7,8], all-normal dispersion [9,10] or noise-like [3,11] regimes, rely on an accurate balance of dispersion inside the cavity to facilitate high pulse energies, wide bandwidths, or both. Recently, a series of commercially available ultra-high numerical aperture fibers (UHNA from Coherent-Nufern) has found use as an intracavity dispersion control tool in ultrafast oscillators working at wavelengths near 2 μm after it had been discovered that these fibers display normal group velocity dispersion (GVD) in the aforementioned spectral region, i.e. they can be used to compensate the anomalous GVD of standard doped and undoped fibers. As UHNAs were originally designed for increasing the coupling efficiency by splicing short sections to waveguides, dispersion compensation, sometimes requiring insertion of fiber sections several meters long, falls outside of the manufacturer's specification. In addition, owing to their small mode areas facilitating nonlinear interactions, supercontinuum generation has been demonstrated in some UHNA fibers [12–15]. In this application the dispersion profile is important as well, as it determines the mechanism of supercontinuum formation. For example, an all-normal dispersion profile facilitates the formation of supercontinuum with high degree of pulse-to-pulse coherence, compressible to short pulse widths using standard ultrafast techniques [16]. Despite these emerging applications, experimental dispersion data on UHNAs is usually not freely accessible from the manufacturer and the values found in the literature are incomplete (i.e. available for a single wavelength only) or inconsistent.

In this paper, we present the results of group-velocity dispersion measurements of Coherent-Nufern UHNA fibers (UHNA1, UHNA3, UHNA4, UHNA7 and PM2000D) over a broad spectral range covering the typically used Er, Tm and Ho emission bands, obtained using the spectral interferometry technique. Standard deviation uncertainty of dispersion resulting from the variability of multiple measurements and the fiber length determination error is calculated and presented in the graphs. The dispersion results are compared with the previous reports found in the literature, and fitting parameters are given to facilitate the straightforward replication of the measured dispersion data.

## 2. EXPERIMENTAL METHOD AND SETUP

For short fiber samples, where direct dispersion measurement methods like the phase shift technique [17] are not viable, spectral interferometry is widely used [18–26]. While interferometer-less schemes do exist [24], the method is usually based on employing a Mach-Zehnder interferometer, with the fiber under test placed in one arm and the other arm serving as the reference (Fig. 1). The interferometric measurement reveals the spectral phase difference between the two interferometer arms (which ideally is only due to the fiber under test), retrieved from the interferogram by i.e. observing equalization wavelength shift for various arm path length differences [19,21], Fourier filtering [22], fringe counting [20,25], nonlinear fit of the interferogram [23] or direct phase extraction [26]. Dispersion is subsequently obtainable by differentiation of the spectral phase.

We employed a fringe-counting spectral interferometric method based on the one reported by Hlubina et al. [25], with modifications. Contrary to the techniques based on Fourier filtering, this approach allows the phase retrieval even when the wide bandwidth of the interferogram and finite resolution of the spectrograph necessitate the inclusion of the equalization wavelength(s) (at which the group delay in both arms is equal) for high visibility of the fringes.

The method depends on assigning the order of the spectral interference fringe versus its wavelength position. A bright fringe on a spectral interferogram satisfies the relation

$$L - l - n(\lambda)z = m\lambda, \qquad (1)$$

where $L$ is the length of the (empty) reference arm of the interferometer, $l$ is the length in air of the test arm in which the fiber to be measured is placed, $n(\lambda)$ is the refractive index of the fiber and $z$ is the fiber length. Counting $i$ bright fringes in the direction of shorter wavelengths yields

$$[L - l - n(\lambda)z]/\lambda = m + i. \qquad (2)$$

By assigning the interference order $i$ to the positions of peaks and valleys on an interferogram one can retrieve the spectral phase shape unambiguously under the following conditions. First, the interference orders shall be numbered such that they go through extrema at equalization wavelengths to reflect the vanishing of the group delay difference. Second, it should be known whether a given equalization wavelength fringe is in the normal or anomalous dispersion region. This information, easily accessed by observing the direction of the spectral shift of the equalization point upon the change of the optical path difference between the interferometer arms, determines whether a given extremum in the fringe order function should be a minimum or a maximum. Only with these conditions met will our constructed interference order function $i(\lambda)$ satisfy Eq. (2) and reflect the shape of the spectral phase. For example, if the shortest equalization wavelength is in the normal dispersion region, then going towards longer wavelengths the fringe order should increase up to the first equalization wavelength, decrease until the second equalization wavelength, then increase once again in the direction of longer wavelengths and so on, depending on the number of equalization points visible in the interferogram.

Subsequent operations enable to retrieve the dispersion parameter $D$ either by direct numerical differentiation [26] or by approximating the refractive index with polynomials. Hlubina et al. [25] used a modified Cauchy dispersion formula in the form

$$n(\lambda) = A_1\lambda^{-4} + A_2\lambda^{-2} + A_3 + A_4\lambda^2 + A_5\lambda^4. \qquad (3)$$

Inserting (3) into (2) yields a function which we found able to fit the interference order function satisfactorily even for relatively complicated phase profiles,

$$i = a_1\lambda^{-5} + a_2\lambda^{-3} + a_3\lambda^{-1} + a_4\lambda + a_5\lambda^3 - m, \qquad (4)$$

where $a_1=-A_1z$, $a_2=-A_2z$, $a_3=L-l-A_3z$, $a_4=-A_4z$, $a_5=-A_5z$ and $m$ are the fitting parameters. Dispersion $D$ can then be calculated as

$$D(\lambda) = \frac{1}{c}\frac{dN(\lambda)}{d\lambda}$$
$$= \frac{1}{c}(-20A_1\lambda^{-5} - 6A_2\lambda^{-3} - 2A_4\lambda - 12A_5\lambda^3), \qquad (5)$$

where $N(\lambda)$ is the group index, given by

$$N(\lambda) = n(\lambda) - \lambda\frac{dn(\lambda)}{d\lambda}. \qquad (6)$$

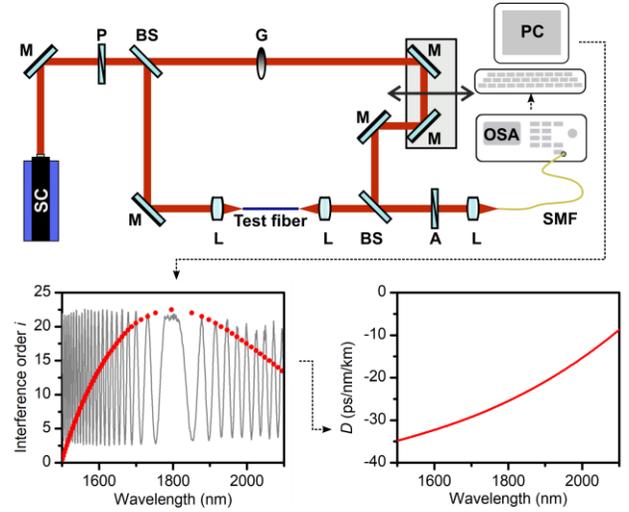

Fig. 1. The experimental setup and workflow for obtaining dispersion by assigning the interference order $i$ to the positions of peaks and valleys of the spectral interferogram. SC – supercontinuum source, M – silver mirror, P – polarizer, BS – beam-splitter, G – gradient neutral-density filter, L – lenses, A – polarization analyzer, SMF – single mode fiber, OSA – optical spectrum analyzer, PC – personal computer.

The experimental setup, shown in Fig. 1, comprised a Mach-Zehnder interferometer with variable path length $L$ in the reference arm. A collimated supercontinuum beam was used as a light source, providing wide bandwidth (450–2400 nm), high brightness and excellent spatial coherence (in contrast, the source offered no pulse to pulse coherence). Matching broadband beam-splitters defined the interferometer arms, while aspheric lenses were used to couple the light to the test fibers and collimate the exiting beam. Care was taken to ensure the proper mode-matching of the input light to the fibers. Coupling was verified by imaging the output face of the fiber on a silicon-chip camera and using a long-pass filter to observe only the wavelengths near 1100 nm, as close to the measured spectral region as possible. The geometry of the setup and the range of the stage used to vary the optical path in the reference arm allowed for fibers 100–200 mm long to be measured, with a typical length of ca. 150 mm. For the measurements of polarization-maintaining (PM) fibers, broadband linear polarizers with parallel axes were placed at the input and output of the interferometer while the fiber could be rotated in its mount to orient the fiber axes. The interferometer output was transferred with a standard single mode patch cord to the optical spectrum analyzer (OSA) capable of recording spectra spanning the range of 1200–2400 nm. It was ensured that the width of the finest fringe on each recorded interferogram was much greater than the spectral resolution of the OSA set at 0.5 nm. Similarly, integration time of 3 ms allowed for multiple supercontinuum pulses to be combined to form the signal at each measurement point.

In the original work [25], the coefficients $A_1$–$A_5$ obtained from the interference order function were used as a first guess for the retrieval of the spectral phase by direct cosine fitting of the interferogram. For our broad spectral range measurements the latter method proved unsuitable. Instead, we globally fitted a set of fringe order functions obtained for a given fiber sample from a number of interferograms with different positions of equalization wavelengths. The sets of 7–17 interferograms allowed for an acceptable fidelity of the results. Moreover, dispersion standard-deviation uncertainties could be established from the standard deviations of the fitted coefficients and the error of determination of the length of the fiber (+/− 1 mm). While the latter is larger than in the original work [25], leveraging the knowledge that for each globally fitted interferogram series the fiber length was exactly the same prevented the overestimation of the uncertainty of dispersion.

Additionally, the results calculated using the fitting parameters were cross-checked with the dispersion obtained by direct numerical differentiation of the spectral phase, in a similar way to Ponzo et al. [26]. While our numerical differentiation operations led to unacceptably noisy curves (not shown), the visual comparison allowed to ensure that no major artifacts resulted from using approximating functions like the one in Eq. (3).

The parasitic dispersion of two thin aspheric lenses used for coupling and collimating the light in the test arm was partially compensated by the neutral-density gradient filter placed in the reference arm for equalization of the intensities of the interfering fields. Nevertheless, the effective dispersion of those elements was determined from the Sellmeier coefficients and subtracted from the experimental results.

include the standard deviations of $A_1$–$A_5$ in Table 1, because the total error of $D$ depends additionally on the error of measuring the length of the fiber. Instead, we indicate the overall standard-deviation uncertainty of the measured dispersion in Figs. 3–7 and include it in the data files (see Supporting Information).

**Table 1. Fitting parameters for reproducing the measured dispersion parameter $D$ using Eq. (5)[a]**

| Fiber | $A_1$ | $A_2$ | $A_4$ | $A_5$ | Applicability range |
|---|---|---|---|---|---|
| UHNA1 | $1.132 \cdot 10^{-6}$ | $-2.059 \cdot 10^{-6}$ | $6.734 \cdot 10^{-6}$ | $-2.239 \cdot 10^{-7}$ | 1200–2400 nm |
| UHNA3 | $-5.855 \cdot 10^{-6}$ | $17.273 \cdot 10^{-6}$ | $5.029 \cdot 10^{-6}$ | $-1.186 \cdot 10^{-7}$ | 1410–2200 nm |
| UHNA4 | $1.982 \cdot 10^{-6}$ | $-5.010 \cdot 10^{-6}$ | $10.433 \cdot 10^{-6}$ | $-2.798 \cdot 10^{-7}$ | 1200–2400 nm |
| UHNA7 | $1.161 \cdot 10^{-9}$ | $1.093 \cdot 10^{-6}$ | $1.443 \cdot 10^{-6}$ | $-3.453 \cdot 10^{-10}$ | 1420–2325 nm |
| PM2000D (slow axis) | $3.766 \cdot 10^{-6}$ | $-9.296 \cdot 10^{-6}$ | $9.677 \cdot 10^{-6}$ | $-2.143 \cdot 10^{-7}$ | 1420–2400 nm |
| PM2000D (fast axis) | $1.102 \cdot 10^{-6}$ | $-3.015 \cdot 10^{-6}$ | $7.752 \cdot 10^{-6}$ | $-1.706 \cdot 10^{-7}$ | 1420–2400 nm |

[a]To obtain $D$ in ps/nm/km, express the speed of light $c$ in km/ps and the wavelength $\lambda$ in µm.

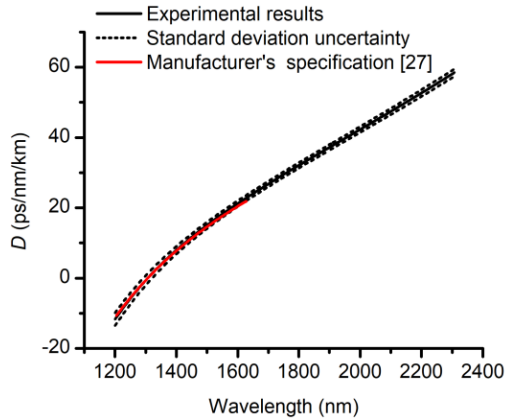

Fig. 2. Measured dispersion $D$ of a standard single-mode fiber SMF-28 Ultra compared with the manufacturer's specification available in 1200–1625 nm region. [27].

The overall performance of the method was verified by measuring the frequency dependence of the dispersion parameter $D$ of a standard single mode fiber SMF-28 Ultra. The results in Fig. 2 show good agreement with the manufacturer's specification [27] in the region 1200–1625 nm, where the latter was available.

## 3. DISPERSION MEASUREMENT RESULTS OF UHNA FIBERS

Coherent-Nufern's ultra-high numerical aperture fibers measured by us comprised UHNA1, UHNA3, UHNA4, UHNA7 and PM2000D, the latter a polarization-maintaining fiber specially designed for ∼2 µm region. In Figs. 3–7 we plot the dispersion spectra of the fibers presented in terms of dispersion parameter $D$ expressed in ps/nm/km, widely used by the fiber community, and as group velocity dispersion $\beta_2$ in ps$^2$/m versus frequency, common for ultrafast science. The latter form especially facilitates judging the amount of higher orders of spectral phase by observing the slope of the curve. The exact wavelength range for which the dispersions were obtained for different fibers depended on the visibility of the fringes. However, the typical erbium, thulium and holmium emission bands were covered in each case. Table 1 shows the fitting coefficients $A_1$–$A_5$ used to generate the plots of the dispersion parameter $D$ shown in Figs. 3–7 and specifies the wavelength range of applicability of the results. We do not

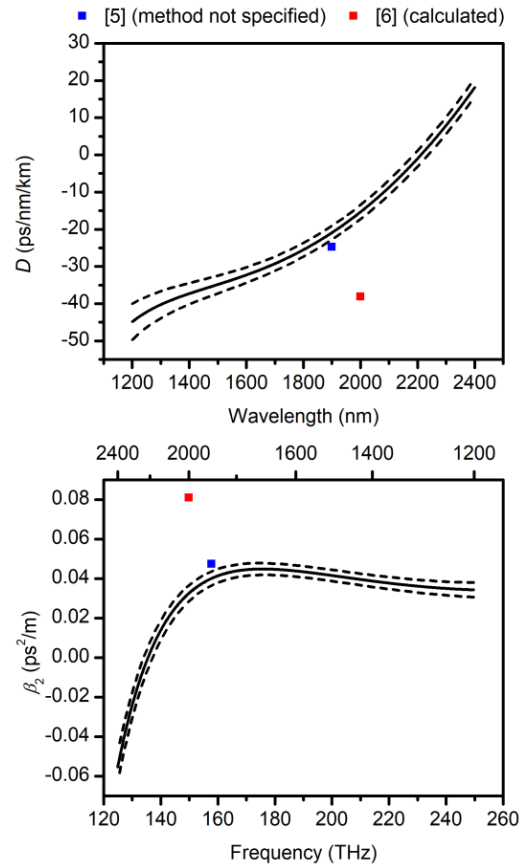

Fig. 3. Dispersion of UHNA1 fiber (black line) presented in terms of the dispersion parameter $D$ (top) and group velocity dispersion $\beta_2$ (bottom). Dashed lines denote the standard-deviation uncertainty ranges. Literature values (squares) are provided for comparison.

Of all the fibers measured, UHNA1 (Fig. 3) is the only one displaying a zero dispersion wavelength within the measured spectral range, found close to 2210 nm. The dispersion is generally low and flat, with $D$ varying only between (−35)–(−15) ps/nm/km in the range 1550–2000 nm. Kadel et al. [6] compared the ability of UHNA1, UHNA4 and UHNA7 to compress the negatively chirped output of a soliton laser operating at 2 µm by observing the resulting pulse autocorrelations. No significant compression was obtained using UHNA1. Indeed, of the three fibers UHNA1 exhibits the lowest normal dispersion value at 2 µm. Kadel et al. also calculated a theoretical value for $\beta_2$ using the formula for step index fibers and the values for the core diameter and numerical aperture provided by the manufacturer. The resulting GVD at 2 µm is more than two times larger than the one measured by us (−0.081 ps$^2$/m [6] vs −0.033 ps$^2$/m). However, the calculated group velocity dispersion values provided by Kadel et al. for UHNA1, UHNA4 and UHNA7 seem regularly overestimated, neither consistent with the authors' own experimental findings nor with our results. We believe this is due to the unknown core material composition, which is not provided by the manufacturer, leading to uncertainties in the calculated dispersion value. A normal dispersion fiber matching the specification of UHNA1 was also employed in the construction of a Thulium fiber laser operating around 1.93 µm [10]. The total cavity dispersion was measured before and after inserting a 6.6 m length of UHNA1, and from the difference the fiber GVD was estimated to $\beta_2$ = 0.036 ps$^2$/m, very close to our measured value $\beta_2$ = 0.038 ps$^2$/m at 1.93 µm. Separately, Wienke et al. [5] estimated the dispersion of the same fiber as 0.0474 ps$^2$/m at 1.9 µm, in good agreement with our value of (0.040+/−0.004) ps$^2$/m.

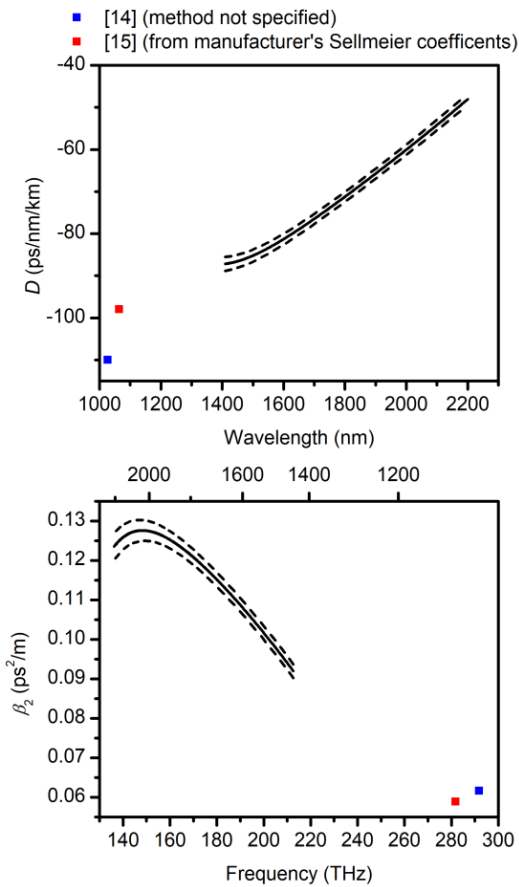

Fig. 4. Dispersion of UHNA3 fiber (black line) presented in terms of the dispersion parameter $D$ (top) and group velocity dispersion $\beta_2$ (bottom) with dashed lines designating the standard-deviation uncertainty ranges. Squares denote the literature values provided for orientation purposes.

UHNA3 displays a much larger normal dispersion than UHNA1, with the dispersion parameter ranging $D$ = (−85) – (−50) ps/nm/km between 1400–2200 nm (Fig. 4). In the frequency domain, the GVD flattens around 2 µm, with zero third order dispersion (TOD) near 2020 nm. We are only aware of two previous dispersion literature values, both of which refer to ~1 µm wavelength [14,15] not accessible with our setup. Nevertheless, we included them into our graphs for orientation purposes.

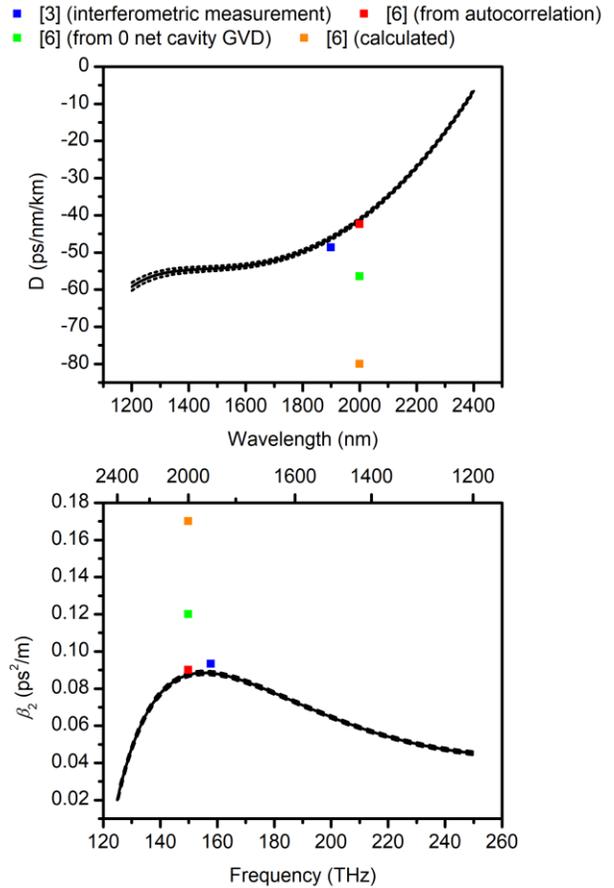

Fig. 5. Dispersion of UHNA4 fiber (black line) presented in terms of the dispersion parameter $D$ (top) and group delay dispersion $\beta_2$ (bottom). Dashed lines denote the standard-deviation uncertainty ranges. Literature values (squares) are included for comparison purposes.

In contrast, several groups reported on the dispersion of UHNA4 around 2 µm (Fig. 5). Spectral interferometry measurements by Wang et al. [3] resulted in $\beta_2$ = 0.09323 ps$^2$/m at 1.9 µm, which is close to our value of (0.0884+/−0.0010) ps$^2$/m. Kadel et al. [6] offer three values of dispersion at 2 µm: calculated (0.17 ps$^2$/m), estimated from autocorrelation pulse width measurements (~0.09 ps$^2$/m) and resulting from the influence of cavity GVD on a Thulium oscillator operation (~0.12 ps$^2$/m). Our results confirm our previous observation that the calculated dispersion values for UHNA fibers yield an overestimation, while the autocorrelation-based measurements show excellent agreement. Wang et al. [3] also report a value of the third order dispersion $\beta_3$ = 1.536·10$^{-4}$ ps$^3$/m, which differs from our result of ca. −2·10$^{-5}$ ps$^3$/m. According to our measurements, the third order zero dispersion wavelength is located at ~1930 nm. At 2 µm, the fiber exhibits a dispersion parameter of $D$ = −40 ps/nm/km, which almost exactly cancels the dispersion of SMF-28 while not adding a significant amount of TOD.

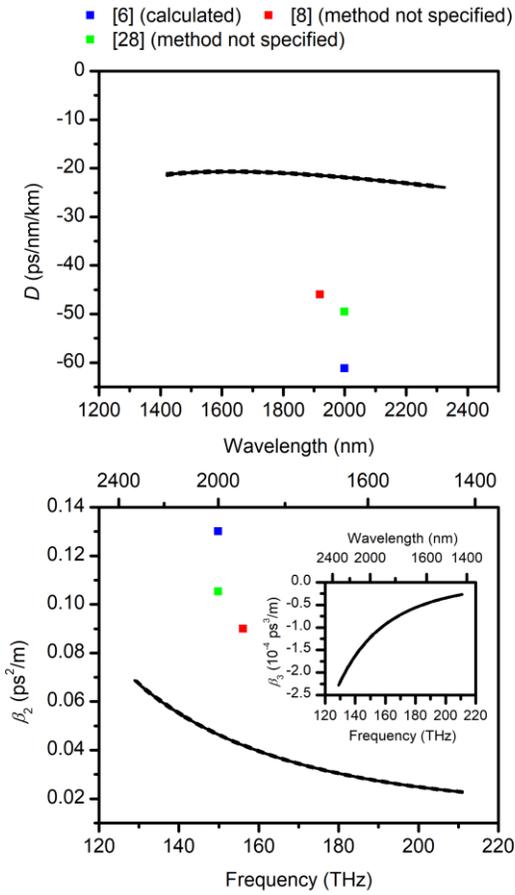

Fig. 6. Dispersion of UHNA7 fiber (black line) presented in terms of the dispersion parameter $D$ (top), group delay dispersion $\beta_2$ (bottom) and third order dispersion $\beta_3$ (bottom inset). Dashed lines denote the standard-deviation uncertainty regions. Literature values (squares) are provided for comparison.

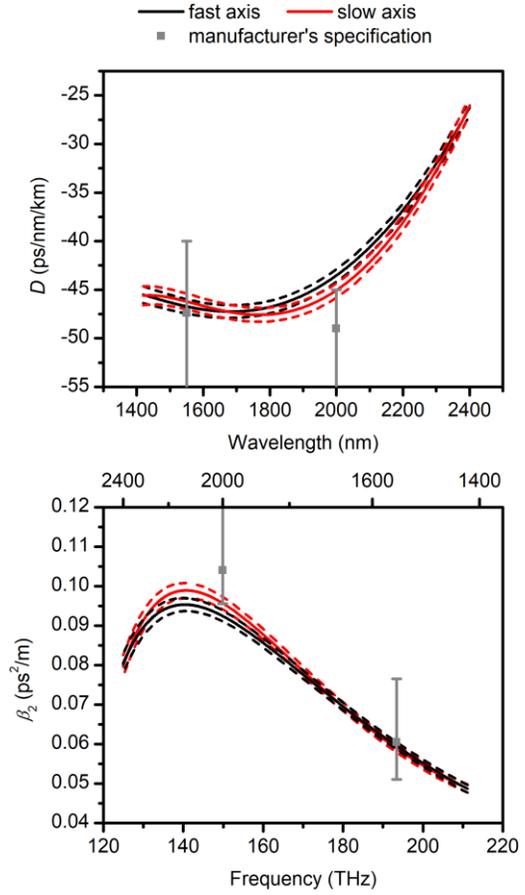

Fig. 7. Dispersion of PM2000D fiber (black line – fast axis, red line – slow axis) given as the dispersion parameter $D$ (top) and group velocity dispersion $\beta_2$ (bottom). Dashed lines denote the standard-deviation uncertainty ranges, and squares designate the manufacturer-specified values.

Fig. 6 shows our measurement results for UHNA7. While exhibiting a low and flat normal dispersion over the entire measurement range, the fiber also displays a significant negative TOD near 2 μm (see inset in Fig. 6). Our measurement yields $\beta_2 = (0.0464+/−0.0004)$ ps$^2$/m and $\beta_3 = −1.2·10^{−4}$ ps$^3$/m at 2000 nm. Therefore, both its 2$^{nd}$ and 3$^{rd}$ dispersion orders are opposite in sign to those of SMF-28, making the fiber potentially useful for compensating higher order dispersion in an ultrafast laser cavity or amplifier. This observation was qualitatively confirmed by the calculations provided by Kadel et al., although the absolute values ($\beta_2 = 0.13$ ps$^2$/m and $\beta_3 = −5.76·10^{−4}$ ps$^3$/m at 2 μm [6]) overestimate the measurements, as was also the case for the other investigated UHNA fibers. Two more dispersion values for UHNA7 can be found in literature ($\beta_2 = 0.09$ ps$^2$/m and $\beta_3 = −3.3·10^{−5}$ ps$^3$/m at 1.92 μm [8]; $D = −49.56$ ps/nm/km ($\beta_2 = \sim 0.1$ ps$^2$/m) near 2 μm [28]), although in both cases the origin of these values is unclear as no measurements or references are provided. Hence, we provide the first detailed dispersion measurement for this fiber, yielding much lower absolute dispersion values than those currently available in literature.

In Fig. 7 we present the measured dispersion characteristics of the PM2000D fiber, a recent addition to Coherent-Nufern's catalogue. It is a panda-type polarization-maintaining fiber with a tailored dispersion designed specifically for the 2 μm wavelength region. To our knowledge, this is currently the only option for polarization-maintaining dispersion control in this wavelength region. Fig. 7 displays the dispersion curves for both polarization axes. For the particular batch tested by us, the manufacturer specifies the dispersion as −47.4 ps/nm/km at 1550 nm and −49 ps/nm/km at 2000 nm. Those values compare well with our measurements. In Fig. 7 we also included the manufacturer's general specification bounds on dispersion. According to the product datasheet, the fiber is designed to exhibit a very low slope of the dispersion parameter $D$, a feature that is well supported by our measurements. $D$ varies only by ~10 ps/nm/km in the range 1420–2100 nm around a central value of −45 ps/nm/km. TOD is flat to slightly negative, the measured zero TOD wavelength is located near 2130 nm. Therefore, the fiber seems indeed well suited for dispersion control in ultrafast fiber laser systems in the 2 μm wavelength region.

## 4. SUMMARY

We measured the group velocity dispersion in a series of commercially-available ultra-high numerical aperture fibers in a wide spectral range which covered the typically used erbium, holmium and thulium emission wavelengths. Widely used for dispersion management in ultrafast fiber laser cavities, the dispersion values of these fibers were previously estimated based on the calculations, with measurement data scarcely available. To address this shortcoming, we employed a spectral interferometry based dispersion measurement method. Validated by measuring a standard SMF-28 Ultra fiber with a good agreement with the catalogue values, it allowed for the robust collection of results in a wide wavelength range. We plotted the results in terms of the dispersion parameter $D$ and the group velocity dispersion $\beta_2$, calculated the standard-deviation uncertainty and offered the tabularized fitting coefficients for all measured dispersion curves. The results were compared with the values available from the literature, with the focus on the 2 µm range. We found that the published calculated dispersion values generally overestimate the measured value, which we believe is due to the unknown core material composition leading to uncertainties in these calculations. However, we observed good agreement with previously published narrowband or single-wavelength experimental values based on autocorrelation, interferometry or total cavity dispersion measurements, where available. We envisage our results to serve as a guideline for the design of ultrafast fiber lasers and amplifiers as they can be used directly as input for the simulation of such systems, as well as for the nonlinear optics community to facilitate the design and simulation of novel supercontinuum light sources.

**Funding Information.** Foundation for Polish Science co-financed by the European Union under the European Regional Development Fund (project number First TEAM/2016-1/1).

## REFERENCES

1. K. Kieu and F. W. Wise, "Soliton Thulium-Doped Fiber Laser With Carbon Nanotube Saturable Absorber," IEEE Photonics Technol. Lett. **21**, 128–130 (2009).
2. R. Kadel and B. R. Washburn, "All-fiber passively mode-locked thulium/holmium laser with two center wavelengths," Appl. Opt. **51**, 6465 (2012).
3. Q. Wang, T. Chen, M. Li, B. Zhang, Y. Lu, and K. P. Chen, "All-fiber ultrafast thulium-doped fiber ring laser with dissipative soliton and noise-like output in normal dispersion by single-wall carbon nanotubes," Appl. Phys. Lett. **103**, 011103 (2013).
4. F. Haxsen, A. Ruehl, M. Engelbrecht, D. Wandt, U. Morgner, and D. Kracht, "Stretched-pulse operation of a thulium-doped fiber laser," Opt. Express **16**, 20471 (2008).
5. A. Wienke, F. Haxsen, D. Wandt, U. Morgner, J. Neumann, and D. Kracht, "Ultrafast, stretched-pulse thulium-doped fiber laser with a fiber-based dispersion management," Opt. Lett. **37**, 2466 (2012).
6. R. Kadel and B. R. Washburn, "Stretched-pulse and solitonic operation of an all-fiber thulium/holmium-doped fiber laser," Appl. Opt. **54**, 746 (2015).
7. H. Li, J. Liu, Z. Cheng, J. Xu, F. Tan, and P. Wang, "Pulse-shaping mechanisms in passively mode-locked thulium-doped fiber lasers," Opt. Express **23**, 6292 (2015).
8. Y. Tang, A. Chong, and F. W. Wise, "Generation of 8 nJ pulses from a normal-dispersion thulium fiber laser," Opt. Lett. **40**, 2361 (2015).
9. H. Liu, "Tm Fiber Laser Mode-Locked At Large Normal Dispersion," in *CLEO:2011 - Laser Applications to Photonic Applications*, OSA Technical Digest (CD) (OSA, 2011), paper CMK1.
10. F. Haxsen, D. Wandt, U. Morgner, J. Neumann, and D. Kracht, "Monotonically chirped pulse evolution in an ultrashort pulse thulium-doped fiber laser," Opt. Lett. **37**, 1014 (2012).
11. G. Sobon, J. Sotor, T. Martynkien, and K. M. Abramski, "Ultra-broadband dissipative soliton and noise-like pulse generation from a normal dispersion mode-locked Tm-doped all-fiber laser," Opt. Express **24**, 6156 (2016).
12. D. L. Marks, A. L. Oldenburg, J. J. Reynolds, and S. A. Boppart, "Study of an ultrahigh-numerical-aperture fiber continuum generation source for optical coherence tomography," Opt. Lett. **27**, 2010 (2002).
13. N. Nishizawa, Y. Chen, P. Hsiung, V. Sharma, T. H. Ko, and J. G. Fujimoto, "All fiber high resolution OCT system using an ultrashort pulse high power fiber laser," in *Conference on Lasers and Electro-Optics/International Quantum Electronics Conference and Photonic Applications Systems Technologies*, Technical Digest (CD) (OSA, 2004), paper CTuBB3.
14. S. R. Domingue and R. A. Bartels, "Overcoming temporal polarization instabilities from the latent birefringence in all-normal dispersion, wave-breaking-extended nonlinear fiber supercontinuum generation," Opt. Express **21**, 13305 (2013).
15. C.-L. Pan, A. Zaytse, Y.-J. You, and C.-H. Li, "Fiber-laser-generated Noise-like Pulses and Their Applications," in *Fiber Laser*, M. C. Paul, ed. (InTech, 2016).
16. A. M. Heidt, A. Hartung, and H. Bartelt, "Generation of Ultrashort and Coherent Supercontinuum Light Pulses in All-Normal Dispersion Fibers," in *The Supercontinuum Laser Source: The Ultimate White Light*, R. R. Alfano, ed. (Springer New York, 2016).
17. B. Costa, D. Mazzoni, M. Puleo, and E. Vezzoni, "Phase shift technique for the measurement of chromatic dispersion in optical fibers using LED's," IEEE J. Quantum Electron. **18**, 1509–1515 (1982).
18. P. A. Merritt, R. P. Tatam, and D. A. Jackson, "Interferometric chromatic dispersion measurements on short lengths of monomode optical fiber," J. Light. Technol. **7**, 703–716 (1989).
19. F. Koch, S. V. Chernikov, and J. R. Taylor, "Dispersion measurement in optical fibres over the entire spectral range from 1.1 µm to 1.7 µm," Opt. Commun. **175**, 209–213 (2000).
20. J. Y. Lee and D. Y. Kim, "Versatile chromatic dispersion measurement of a single mode fiber using spectral white light interferometry," Opt. Express **14**, 11608 (2006).
21. P. Hlubina, M. Szpulak, D. Ciprian, T. Martynkien, and W. Urbanczyk, "Measurement of the group dispersion of the fundamental mode of holey fiber by white-light spectral interferometry," Opt. Express **15**, 11073 (2007).
22. P. Hlubina, J. Luňáček, D. Ciprian, and R. Chlebus, "Windowed Fourier transform applied in the wavelength domain to process the spectral interference signals," Opt. Commun. **281**, 2349–2354 (2008).
23. T. M. Kardaś and C. Radzewicz, "Broadband near-infrared fibers dispersion measurement using white-light spectral interferometry," Opt. Commun. **282**, 4361–4365 (2009).
24. N. K. Berger, B. Levit, and B. Fischer, "Measurement of fiber chromatic dispersion using spectral interferometry with modulation of dispersed laser pulses," Opt. Commun. **283**, 3953–3956 (2010).
25. P. Hlubina, M. Kadulová, and D. Ciprian, "Spectral interferometry-based chromatic dispersion measurement of fibre including the zero-dispersion wavelength," J. Eur. Opt. Soc. Rapid Publ. **7**, (2012).
26. G. M. Ponzo, M. N. Petrovich, X. Feng, P. Horak, F. Poletti, P. Petropoulos, and D. J. Richardson, "Fast and broadband fiber dispersion measurement with dense wavelength sampling," Opt. Express **22**, 943 (2014).
27. Corning Incorporated, "Corning SMF-28 Ultra Optical Fiber Product Information," https://www.corning.com/media/worldwide/coc/documents/Fiber_11-14.pdf.
28. J. Luo, B. Sun, J. Liu, Z. Yan, N. Li, E. L. Tan, Q. Wang, and X. Yu, "Mid-IR supercontinuum pumped by femtosecond pulses from thulium doped all-fiber amplifier," Opt. Express **24**, 13939 (2016).